\documentclass[prd,aps,nofootinbib,twocolumn]{revtex4-1}
\usepackage{amsmath,amssymb,amsfonts,color,graphicx,graphics,latexsym,placeins,epsfig,multirow}
\usepackage{array}
\newcolumntype{P}[1]{>{\centering\arraybackslash}p{#1}}
\usepackage{footmisc}
\usepackage[compatibility=false]{caption}
\usepackage{subcaption}
\usepackage{comment}
\usepackage{bbold}
\usepackage{enumitem}
\setdescription{leftmargin=8pt}
\usepackage{bbold}
\usepackage{hyperref}
\hypersetup{
	colorlinks=true,       % false: boxed links; true: colored links
	linkcolor=blue,          % color of internal links
	citecolor=blue,        % color of links to bibliography
}

\usepackage{natbib}
\setcitestyle{square, comma, numbers, sort&compress}
\usepackage{float}

\usepackage{amsmath,amssymb}
\newcommand{\ba}{\begin{align}}
\newcommand{\ea}{\end{align}}

\def\3nab{\tilde{\nabla}}

\def\be {\begin{equation}}
\def\ee {\end{equation}}
\def\ba {\begin{eqnarray}}
\def\ea {\end{eqnarray}}

\newcommand{\sfr}[2]
{{\textstyle\frac{#1}{#2}}}

\newcommand{\barray}{\begin{array}}
\newcommand{\earray}{\end{array}}

\newcommand{\bea}{\begin{eqnarray}}
\newcommand{\eea}{\end{eqnarray}}

\newcommand{\nn}{{\nonumber}}

\usepackage[dvipsnames]{xcolor}

\begin{document}

%%%%%%%%%%%%%%% AUTHOR'S NAMES AND AFFILIATIONS %%%%%%%%%%%%%%%%%%

\title{\Large \bf Gravitational memory signal from neutrino self-interactions in
supernova}

%\title{\Large \bf A master equation for Gravitational Wave Memory in Cosmology}

%\title{\Large \bf Distinguishing/ probing accelerated and decelerated phases of the Universe using GW memory}

%\title{\Large \bf Grav memory signatures in cosmological era transitions}

\author{\sf Soumya Bhattacharya$^1$} \email{soumya557@gmail.com}

\author{\sf Debanjan Bose$^2$} \email{debaice@gmail.com}

\author{\sf Indranil Chakraborty$^3$}
\email{indranil.phy@iitb.ac.in (Corresponding author)}

\author{\sf Arpan Hait$^4$} \email{arpan20@iitk.ac.in}

 \author{\sf Subhendra Mohanty$^4$} \email{mohantys@iitk.ac.in}

\vspace{0.1in}

 \affiliation{$^1$ S. N. Bose National Centre for Basic Sciences, Kolkata 700106, India}

\affiliation{$^2$Department of Physics,Central University of Kashmir, Ganderbal 191131, India}
 
 \affiliation{$^3$Department of Physics,  Indian Institute of Technology Bombay, Mumbai 400076, India}

 \affiliation{$^4$ Department of Physics, Indian Institute of Technology Kanpur, Kanpur 208016, India}
 
	%%%%%%%%%%%%%%%%%%%%
	
	\begin{abstract}
     	Neutrinos with large self interactions arising from exchange of light scalars or vectors with mass $M_\phi\simeq 10{\rm MeV}$ can play a useful role in cosmology for structure formation and solving the Hubble tension. It has been proposed that large self interactions of neutrinos may change the observed properties of supernova like the neutrino luminosity or the duration of the neutrino burst. In this letter, we compare the gravitational wave memory signal  arising from supernova neutrinos. Our results reveal that memory signal for self-interacting neutrinos are weaker than free-streaming neutrinos in the high frequency range. Implications for detecting and differentiating between such signals for planned space-borne detectors, DECIGO and BBO, are also discussed.
				\end{abstract}

	%%%%
	
	\maketitle

\noindent \underline{\emph{Introduction:}} Even before the observation of neutrinos from the supernova SN1987A \cite{Arnett:1989tnf,Chauhan:2021}, it has been recognised that core-collapse supernova, which produces a high density of neutrinos, would be ideal for the study of neutrino self-interactions ($\nu$-SI) \cite{DICUS:1983, Abbott:2016, Manohar, Berenzhiani, Choi, masso, DICUS:1989, Aharonov, Fuller, BEREZHIANI:1989, Farzan:2002, Serpico, Heurtier:2016, Dighe, Shalgar, Cerdeno:2023kqo, Chang:2022aas, Raffelt-1, Brdar:2023tmi,Fiorillo:2023,Fiorillo:2023cas,Akita:2022etk}. Among various properties of neutrino, $\nu$-SI  mediated by light scalars or vectors are of interest  for both laboratory and cosmological applications \cite{Berryman:2022}. $\nu$-SI with scalar or vector mediators as light as $M_\phi\sim 10$ MeV can have useful applications in cosmology, and are allowed by neutrino experiments and CMB observations \cite{Lancaster:2017ksf,Oldengott:2017fhy,Kreisch:2019yzn}. SI of massive neutrinos reduces the free-streaming length and can be detected  in CMB and large-scale structure observation. Moreover, this leads to a modification in the allowed ($n_s$, $r$) parameter space of inflationary models \cite{Barenboim:2019tux,Mazumdar:2019tbm}. Flavour specific $\nu$-SI can alleviate the $H_0$ tension \cite{Blinov:2019gcj,Blinov:2020hmc,He:2020zns,Mazumdar:2020ibx}, while being allowed by collider constraints \cite{Brdar:2020nbj}. High energy neutrinos can be scattered or absorbed by the cosmic neutrino background and produce a dip in the observations of neutrino spectrum at IceCube which will be a signal of $\nu$-SIs \cite{Ng:2014pca,DiFranzo:2015qea,Shoemaker:2015qul,Bustamante:2020mep}.
	
On the question of the effect of $\nu$-SI on the neutrino signal from core-collapse supernova there is no universal consensus. In the particle picture, one assumes that SIs would lead to successive scatterings of the emitted neutrinos, which in case of large $\nu$-SI, could lead to neutrino trapping and a reduction in the observed flux \cite{Manohar}. In \cite{DICUS:1989}, it was shown that interacting neutrinos act as a fluid with sub-luminal velocities for a certain distance, beyond which they free-stream at luminal velocities. Taking motivation from this, the authors in Ref.~\cite{Chang:2022aas} have studied the effect of $\nu$-SI on supernova neutrino signal.  They argued that for large SIs in a {\em burst model} the duration of the neutrino signal is prolonged compared to the standard neutrino interaction. Following this study, Fiorillo et al. \cite{Fiorillo:2023,Fiorillo:2023cas} have argued that the more likely scenario for $\nu$-SI in supernova is a steady emission of neutrino from the proto-neutron star (PNS) surface which propagates as a pressure wave with {\em sonic velocity} $\sim 1/\sqrt{3}$ close to the PNS surface and increases to unity at the point where the neutrinos start free-streaming. In this {\em steady wind} model, despite the presence of SIs, the observable neutrino signal ({\em i.e.} the neutrino flux at the detector and the duration of the neutrino signal) remains close to the case of SM-neutrinos, signifying that neutrino observations from supernova do not have a robust signature of SIs in neutrinos.

%% NEED TO CHANGE THIS %%%
 
 In order to check for the signatures of $\nu$-SI arising in such scenarios, we examine the gravitational waves sourced by the supernova neutrinos. It has long  been pointed out that null-fluids like gravitons emitted from inspiraling binaries \cite{Will:1996zj,Hait:2022ukn}  or neutrinos from supernova \cite{Kolb:1987,Epstein:1978,Turner:1978}  can be a source of gravitational waves and gives rise to a step-function like {\em memory effect} in the observed GW signal (For more recent works look at \cite{Mukhopadhyay:2021zbt,Li:2017,Yakunin:2015}).  We compute the memory signal generated by self-interacting neutrinos after they are emitted  from the PNS surface. In a region between the radii $R_{s}< r< R_{fs}$ where $R_{s}\sim 10$ km is the PNS radius and $R_{fs}$, known as the free-streaming radius, is the value at which the neutrinos do not suffer any scattering and begin free-streaming. We find that when neutrinos deviate from luminal velocities in the region $R_s < r <R_{fs}$, the gravitational memory signal they produce become {\em significantly weaker, yet detectable} from the case when there is no $\nu$-SI. Thus, our letter serves as a {\em proof-of-principle} for probing $\nu$-SI using gravitational wave memory. {  Moreover, we comment on how the free-streaming radius is dependent on the strength of $\nu$-SI.} \\

\newpage

\noindent \underline{\emph{$\nu$-SI in supernova:}}
We study neutrino self-interactions of Majorana neutrinos of the form
	\be
	{\cal L}=-\frac{1}{2} g \nu^T \nu \phi, 
	\ee 
	which can arise in Majoron models \cite{Gelmini:1980,Chikashige:1980,Konoplich:1988mj} in which lepton number is broken spontaneously and where $\phi$ is the pseudo Nambu-Goldstone boson   with a small mass which can be as low as $M_\phi \simeq 10 $ MeV. The scalar exchange gives rise to a four-Fermi interaction at a scale lower than $M_\phi$ with a coupling $G^\prime=g^2 /M_\phi^2$ which can be larger than the Fermi constant of weak interaction $G_F$.  {  Initially neutrinos emitted from the PNS surface have an average number density
 $n_\nu^I =10^{36}\, {\rm cm^{-3}}$,  and SI cross-section of
 \begin{equation} \label{eq:SI_cross-section}
     \sigma_{\nu \nu}=g^4/(4\pi M_\phi^2)=\frac{1}{4\pi}(G^\prime)^2\, M_\phi^2.
 \end{equation}
 
However, this density falls as the neutrino travels outwards from the PNS.}
{  The mean-free path for the $\nu$-SI becomes,
	\begin{align} \label{eq:MFP}
	\lambda_{\rm mfp}^I=\frac{1}{n_\nu^I\, \sigma_{\nu \nu}}=16.6\, {\rm km} \left(\frac{3.73 \times 10^{-5}}{g}\right)^4   \left(\frac{M_\phi}{10 \,{\rm MeV}}\right)^2.
	\end{align}
 Near the PNS, the $\nu$-SI coupling $g= 3.73\times 10^{-5}$. The self-interaction strength corresponding to coupling $g \sim 10^{-5}$ with $M_\phi\sim 10\, \text{MeV}$ is similar to the weak-interaction strength, $G_F$.}

After coming out of PNS, the neutrino-fluid undergoes multiple scattering close to the surface of the PNS and only free-streams at a distance $R_{fs} > R_{s}= {\rm 10\, km}$ where the density drops sufficiently so that the optical depth becomes less than unity. The radial distance $r$ where the neutrinos start can be determined from the optical depth at distance $r$
{  \begin{equation} \label{eq:optical-depth}
\tau(r) = -\int^{r}_{\infty} dr\, n_\nu (r)  \sigma_{\nu \nu}\,.	
\end{equation}}

The distance where the optical depth $\tau(r =R_{fs})=1 $ defines the free-streaming radius. The self-interacting neutrinos from supernova  behave as a fluid with radial velocity which varies with distance from the PNS (proto-neutron star) radius $r=R_{s}\simeq 10 {\rm km}$ to $r <R_{fs}$ \cite{Chang:2022aas,Fiorillo:2023}.
	In this diffusion zone, the neutrino fluid has a velocity $v(r< R_{fs})\sim 1/\sqrt{3}$. At $r\geq R_{fs}$ the neutrinos free-steam with the speed of light, $v(r\geq R_{fs})=1$ as described in Ref.~\cite{Chang:2022aas}.  

   The introduction of a new length scale in the problem provides clues on the self-interaction strength of the neutrino. {   We assume the neutrino density follows the power law,
\begin{equation} \label{eq:power-law}
    n_\nu (r) = n_\nu^I\, \bigg(\frac{r}{R_S} \bigg)^{-\beta}.
    \end{equation} 
Here $\beta$ denotes the rapidity with which the density falls and depends on the details of the explosion. Plugging the neutrino density in Eq.(\ref{eq:optical-depth}) we get
\begin{equation} \label{eq:optical-depth-relation}
    \tau(r)= \frac{n_\nu^I\, \sigma_{\nu\nu}\, R_s}{\beta-1} \bigg(\frac{r}{R_s}\bigg)^{1-\beta}=\frac{R_s}{\lambda_{\rm mfp}^I(\beta-1)} \bigg(\frac{r}{R_s}\bigg)^{1-\beta}
\end{equation}
 Substituting from Eq.(\ref{eq:MFP}) the expression of $\lambda_{\text{mfp}}^I$in Eq.(\ref{eq:optical-depth-relation}) we find that for $\beta=2$, 
 \begin{eqnarray}
     R_{fs}=100\, \text{km} \, \, \bigg( \frac{g}{7.53\times10^{-5}} \bigg)^4 \, \bigg( \frac{10 \,\text{MeV}}{M_\phi} \bigg)^2, \noindent\\
      R_{fs}=10^5\, \text{km} \, \, \bigg( \frac{g}{4.23\times10^{-4}} \bigg)^4 \, \bigg( \frac{10 \,\text{MeV}}{M_\phi} \bigg)^2 \noindent.
 \end{eqnarray}
For $\beta=3$, 
 \begin{eqnarray}
     R_{fs}=100\, \text{km} \, \, \bigg( \frac{g}{1.59\times10^{-4}} \bigg)^2 \, \bigg( \frac{10 \,\text{MeV}}{M_\phi} \bigg),\noindent\\
     R_{fs}=10^5\, \text{km} \, \, \bigg( \frac{g}{5.03\times10^{-3}} \bigg)^2 \, \bigg( \frac{10 \,\text{MeV}}{M_\phi} \bigg).
 \end{eqnarray}
} 
In the present letter, we have worked with two values of $R_{fs}$, $100$ km and $10^5$ km, {\em i.e.} small and large diffusion regions. {  When $R_{fs}=100 \, \text{km}$, the coupling $g=7.53\times 10^{-5}$ and $1.59\times 10^{-4}$, for $\beta=2\, \text{and}\, 3$, respectively. For $R_{fs}=10^5 \, \text{km}$, the coupling $g=4.23\times 10^{-4}$ and $5.03 \times10^{-3}$, for $\beta=2\, \text{and}\, 3$, respectively. Thus, we find that with rise in the value of SI coupling $g$, and thereby $G^\prime$, $R_{fs}$ also increases. }   
 
 We will see later in this letter how gravitational memory is dependent on the extent of the diffusion zone and thereby can probe the strength of $\nu$-SI.\\ % Modify this part
 
 \noindent \emph{\underline{Gravitational waves from self-interacting neutrinos:}} We describe the formalism in which GW radiation is produced due to supernova neutrinos.  This radiation carries energy of the neutrinos which contributes to the GW memory signal. Initially, in the diffusion region $R_{s} < r < R_{fs}$,  the neutrino is described by a relativistic fluid moving with a sub-luminal velocity, $v \sim 1/\sqrt{3}$ \cite{Chang:2022aas}. The corresponding gravitational perturbation for such a relativistic fluid is derived in Appendix \ref{app:GW_formula}. We rewrite the expression for convenience,
 \vspace{-0.08in}
\begin{align}\label{hijtgw-3a}
h^{ij}(t, \vec x)
&=\frac{4G}{r} \int  dt^\prime \,d\Omega^\prime \delta \left(t^\prime- (t-r) \right) 
\nonumber
\\
& \hskip 0.5cm \int^{R_{fs}}_{R_s} d r^\prime {r^\prime}^2  \epsilon(\vec x^\prime, t^\prime)\, \frac{v^i(\vec x^\prime, t^\prime) v^j(\vec x^\prime, t^\prime)}{ 1-\vec N \cdot \vec v (\vec x^\prime, t^\prime) }.
\end{align}

After emerging from the diffusion region, neutrinos free-stream ($v=1$) in the region $r> R_{fs}$. The radiated gravitational waves as given by the null-fluid expression is also derived in the Appendix \ref{app:GW_formula}. It reads,
\begin{align}\label{hijtgw-4a}
h^{ij}(t, \vec x)
&=\frac{4G}{r} \int dt^\prime \,d\Omega^\prime \delta \left(t^\prime- (t-r) \right)  
\nonumber
\\
& \hskip 0.5cm \int_{R_{fs}}^{\infty} d r^\prime  \, {r^\prime}^2\epsilon(\vec x^\prime, t^\prime)\, \frac{{n^\prime}^i {n^\prime}^j}{ 1-\vec N \cdot \vec  {n^\prime}}.
\end{align}
Therefore, the total GW radiation from supernova $\nu$-SI is the sum of Eqs.~\eqref{hijtgw-3a} and  \eqref{hijtgw-4a},
\begin{widetext}
    \begin{align}\label{hijtgw-5a}
	h^{ij}(t, \vec x)
=\frac{4G}{r} \int dt^\prime \delta \left(t^\prime- (t-r) \right) \,d\Omega^\prime 
\bigg\{  \int^{R_{fs}}_{R_s} {r^\prime}^2 d r^\prime \epsilon(\vec x^\prime, t^\prime) \frac{v^i(\vec x^\prime, t^\prime) v^j(\vec x^\prime, t^\prime)}{ 1-\vec N \cdot \vec v (\vec x^\prime, t^\prime) } + 
\int_{R_{fs}}^{\infty} {r^\prime}^2 d r^\prime \epsilon(\vec x^\prime, t^\prime) \frac{{n^\prime}^i {n^\prime}^j}{ 1-\vec N \cdot \vec  {n^\prime}}\bigg\}.
\end{align}
\end{widetext}
In the limit of weak $\nu$-SI, we have $R_{fs} \rightarrow R_s$. Here we obtain the standard neutrino memory signal as in  Ref.~\cite{Mukhopadhyay:2021zbt}. The flux density of the neutrinos radiated from the PNS surface can be written in terms of the neutrino luminosity as,
	\begin{equation} \label{eq:energy_density}
	\epsilon(\vec x^\prime, t^\prime) =\frac{L_{\nu_i} (t^\prime, r^\prime)} {4 \pi {r^\prime}^2}\alpha(\theta^\prime, \phi^\prime),
	\end{equation}
	where $L_{\nu_i} (t^\prime, r^\prime)$ is the neutrino luminosity and $\alpha(\theta^\prime, \phi^\prime)$ is the anisotropy parameter \footnote{Note in this work we have only considered a model which has time-independent anisotropy. Such models have been discussed in \cite{Mukhopadhyay:2021zbt} as the {\em wlCA} model. While this model is simplistic, it essentially captures the basic physics of both $\nu$-SI and gravitational memory. Furthermore, there have been works on such constant time anisotropy for Gamma Ray Bursts \cite{GRB_model}.} which describes the angular asymmetry in the neutrino luminosity due to non-spherical collapse of the supernova.
{  In all our calculations henceforth, we have used the same anisotropy parameter $\alpha(\theta^\prime,\phi^\prime) =\alpha\, \cos^2\phi^\prime$. Here, $\alpha$ denotes the strength of anisotropy and the azimuthal part provides the profile of the burst. 
If there is no anisotropy in the neutrino burst, there is no gravitational radiation. For a radial burst, there will be no such memory signal. Later, in the paper we will show why non-radial anisotropy parameter is essential for producing the required memory signal.}
 
The neutrino luminosity is given by \cite{Ko:2022},
\begin{equation} \label{eq:luminosity}
L_{\nu_i} (t^\prime, r^\prime)=\frac{1}{6} \frac{E_b}{\tau_\nu}\exp\left(- \frac{v_r t^\prime-r^\prime}{\tau_\nu}\right) \Theta(v_r t^\prime-r^\prime),
\end{equation}
where $E_b=3 \times 10^{53}$ ergs is the total energy in the explosion, the decay time of the luminosity is $\tau_\nu= 3\, {\rm sec}$, and $v_r$ is the radial velocity of the neutrino fluid. 

In order to clearly outline the role of $\nu$-SI we first evaluate Eq.~\eqref{hijtgw-5a} in two limiting cases. In the limit of weak SIs, the free-streaming of the neutrinos starts from PNS at $r=R_s$. In this limit the gravitational wave signal reduces to the standard result
\begin{align}\label{hijtgw-6a}
    h^{ij}(t, \vec x)
&=\frac{4G}{r} \int dt^\prime \delta \left(t^\prime- (t-r) \right)  \int_{R_{s}}^{\infty} {r^\prime}^2 d r^\prime \epsilon(\vec x^\prime, t^\prime) 
\nonumber
\\
& \hskip 1cm \times\int\, d\Omega^\prime \frac{{n^\prime}^i {n^\prime}^j}{ 1-\vec N \cdot \vec  {n^\prime}}.
\end{align}
On the other hand, for strong self interactions the free-streaming will occur at a distance $R_{fs} \gg R_s$ ({\em i.e.} $R_{fs}\to \infty$) and the signal given by Eq.~\eqref{hijtgw-5a} will be dominated by the first term
\begin{align}\label{hijtgw-7a}
h^{ij}(t, \vec x)
=\frac{4G}{r} \int dt^\prime \delta \left(t^\prime- (t-r) \right) \,d\Omega^\prime   
\nonumber
\\
\times \int^{\infty}_{R_s} {r^\prime}^2 d r^\prime \epsilon(\vec x^\prime, t^\prime) \frac{v^i(\vec x^\prime, t^\prime) v^j(\vec x^\prime, t^\prime)}{ 1-\vec N \cdot \vec v (\vec x^\prime, t^\prime) } .
\end{align}	
Thus, at first, we compute the memory signal entirely without any $\nu$-SI, then we find it for strong $\nu$-SI given in Eq.(\ref{hijtgw-7a}).  Finally, we evaluate the memory signal given in Eq.(\ref{hijtgw-5a}). A comparison of these three scenarios would provide pointers in leveraging gravitational memory as a probe for $\nu$-SI. 
\begin{figure}[htbp] \label{fig:orientation}
    \centering
    \includegraphics[scale=0.2]{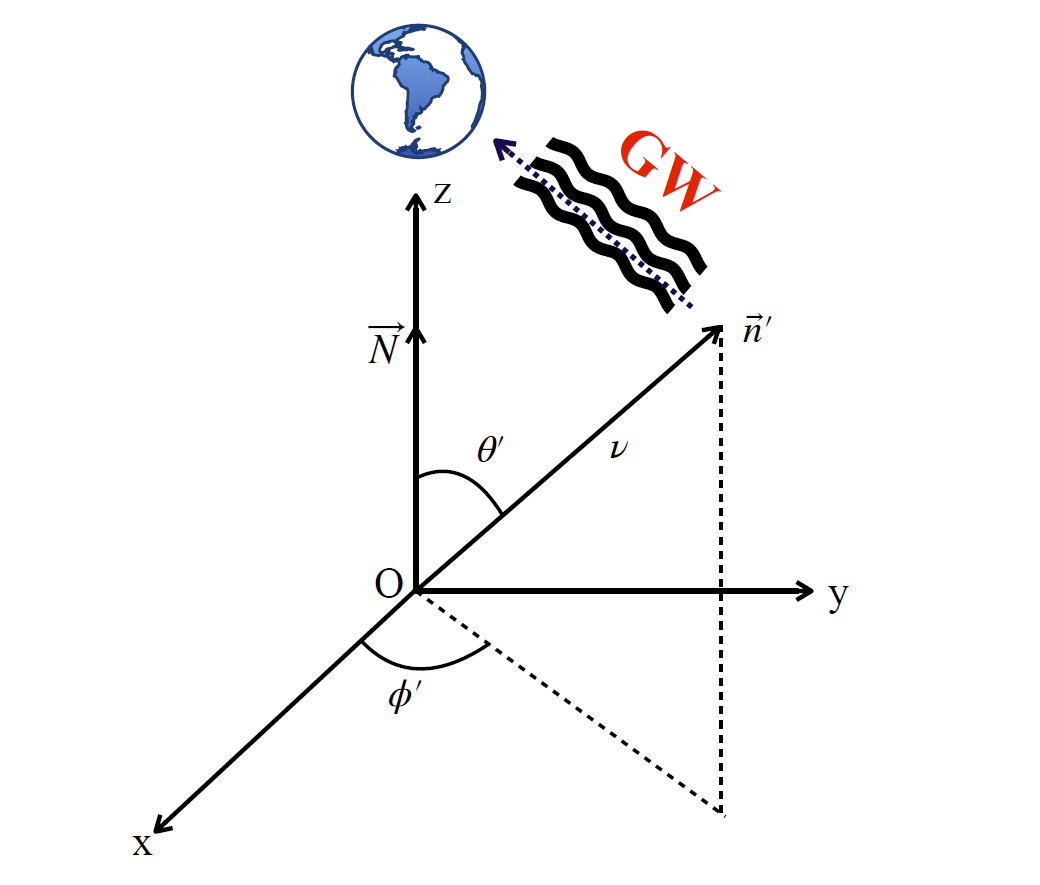}
    \caption{  The figure depicts the position of the detector {\em vis-a-vis} the anisotropic neutrino emission. In our setup, we have considered the $z$-direction to be along the detector. $\theta$ denotes the angle between the neutrino burst ($\vec{n^\prime}$ for null fluid and $\vec{v^\prime}$ for relativistic fluids) and the detector ($\Vec{N}$).}
    \label{fig:observer}
\end{figure}\\

\noindent \underline{\em Time domain Gravitational memory waveforms:} We provide, respectively, the expressions for the memory integral for the three scenarios described previously.
 
%\noindent The gravitational memory integral formula was first given by Thorne \cite{Thorne:1991}. It shows how the integral of the stress energy for gravitational waves (or primary hard gravitons) leads to a memory. In our case, the source is comprised of the supernova neutrinos. To this end, we employ Eq.(\ref{eq:h_ij}) in the TT gauge to find the memory integral. Furthermore, we need to assume anisotropy in the neutrino distribution. This asymmetric emission of neutrinos leading to memory excluding self-interactions has been studied extensively in the literature.

 %In the present article, memory waveforms in three distinct cases are studied-- (i) when neutrino free streams completely ({\em i.e.} no $\nu$-SI) with speed equal to unity, (ii) when the $\nu$-SI is present throughout ($v=\sqrt{3}$), (iii) when the neutrino has a mixed propagation mode. In this scenario, the neutrino travels from $R_s$ to $R_{fs}$ with $v=1/\sqrt{3}$ and for $r>R_{fs}$, it free streams. 

 In the limit of weak $\nu$-SI, we find that $R_{fs}=R_s$. There is no diffusion region present in the neutrino propagation. Thus, the Majorana neutrinos free-stream after leaving the PNS surface. Assuming the velocity to be unity, the expression for the memory integral is detailed below.
\begin{align}\label{eq:mem_SI_low}
	[h_{ij}^{\text{mem}}(t, \vec x)]^{TT} 
&=\frac{4G}{r} \int dt^\prime \delta \left(t^\prime- (t-r) \right) \,d\Omega^\prime 
\nonumber 
\\
& \times \int_{R_{s}}^{\infty} {r^\prime}^2 d r^\prime \epsilon(\vec x^\prime, t^\prime) \bigg[\frac{{n^\prime}^i {n^\prime}^j}{ 1-\vec N \cdot \vec  {n^\prime}}\bigg]^{TT}.
\end{align}
Substituting the Eqs.~\eqref{eq:energy_density}, and~\eqref{eq:luminosity} in Eq.~\eqref{eq:mem_SI_low} we find,
\begin{equation}
    [h_{ij}^{\text{mem}}(t, \vec x)]^{TT} = \frac{G}{6\pi r} E_b \, \bigg(1-\mathrm{Exp}\bigg[-\frac{(t-r)-R_s}{\tau_\nu}\bigg]\bigg) \, \mathcal{A}_{ij}.
\end{equation}
Here $\mathcal{A}_{ij}$ is basically the angular integral. Depending on the two polarizations of the GW radiation, their expressions become,
\begin{align}  \label{eq:angular-integrals-1}
&\mathcal{A}_{ij}=\int d\Omega^\prime \alpha(\theta^\prime,\phi^\prime)\bigg[\frac{{n^\prime}^i {n^\prime}^j}{ 1-\vec N \cdot \vec  {n^\prime}}\bigg]^{TT}
\nonumber \\
  &  \mathcal{A}_{+}= \alpha \, \pi, \qquad\mathcal{A}_{\times}=0.
\end{align}
  The azimuthal dependence of the anisotropy parameter [$\alpha(\theta^\prime,\phi^\prime)= \alpha \cos^2\phi^\prime$] is crucial as, otherwise, the integrals vanish. For more realistic models involving accretion, anisotropy will have non-trivial dependence on $u,\theta^\prime,\phi^\prime$ with inputs from numerical supernova simulations \cite{Kotake:2009, Muller:2011, Vartanyan:2020, Mukhopadhyay:2021zbt}. However, the point of choosing such a simple form is to establish the ubiquitous nature of memory for any simple asymmetric profile. It also provides analytical solvability and gives closed-form expressions for time-domain and frequency-domain memory signals. {  In addition, anisotropy can exhibit spatial dependence. For certain profiles, the memory signal obtained can show a significantly different contribution from the second term in Eq.(\ref{hijtgw-5a}). We provide comments on the qualitative features of the memory signal involving three different kinds of radial anisotropic profile in Appendix \ref{app:rad-ani}.} 

{  We find no memory strain in the cross-polarization as given in Eq.~\eqref{eq:angular-integrals-1}. This is dependent on the choice of the direction of the observer which is specified in Fig.~\eqref{fig:observer}.} Hence,the memory strain in plus polarization becomes
\begin{equation} \label{eq:time_domain_memory_without_SI}
     [h_{+}^{\text{mem}}(t, \vec x)]^{TT} = \frac{G\alpha}{6 r} E_b \, \bigg(1-\mathrm{Exp}\bigg[-\frac{u-R_s}{\tau_\nu}\bigg]\bigg) \Theta (u-R_s).
\end{equation}
We have set $t-r=u$, and at large retarded time $u\to+\infty$, the memory strain yields $h^{\text{mem}}\sim 10^{-21}$.  Our estimates are in agreement with the results found in Ref.~\cite{Mukhopadhyay:2021zbt}.

In the opposite regime, {\em i.e.,} when $\nu$-SI is quite high, the neutrinos after coming out of the PNS encounter a dense environment of other neutrinos. In this approximation, we take the radial velocity of neutrinos to be $v_r=\frac{1}{\sqrt{3}}$\cite{Fiorillo:2023}. 
\begin{eqnarray}\label{eq:mem_SI_high}
&&[h_{ij}^{\text{mem}}(t, \vec x)]^{TT}
= \frac{4G}{r} \int dt^\prime \delta \left(t^\prime- (t-r) \right) \,d\Omega^\prime  
\nonumber
\\
&& \times \int^{\infty}_{R_s} {r^\prime}^2 d r^\prime \epsilon(\vec x^\prime, t^\prime) \frac{v^i(\vec x^\prime, t^\prime) v^j(\vec x^\prime, t^\prime)}{ 1-\vec N \cdot \vec v (\vec x^\prime, t^\prime)}.
\end{eqnarray}
Since the radial velocity $v_r=\frac{1}{\sqrt{3}}$, we take $v^i(\vec{x^\prime}, t^\prime)=v_r \, {n^\prime}^i$. Taking the same anisotropy parameter from the previous calculation, we find the memory strain to be,
\begin{equation}
    [h_{ij}^{\text{mem}}(t, \vec x)]^{TT} = \frac{G}{18\pi r}\,  E_b \, \mathcal{B}_{ij} (1-e^{-\frac{(t-r)-\sqrt{3}R_s}{\sqrt{3}\tau_\nu}}) .
\end{equation}
The modified angular integral in this case and its solution for the two polarizations are given below.
\begin{align} \label{plus_high_SI}
&\mathcal{B}_{ij}=\int d\Omega^\prime \alpha(\theta^\prime,\phi^\prime)\bigg[\frac{{n^\prime}^i {n^\prime}^j}{ 1-v_r\, (\vec N \cdot \vec  {n^\prime})}\bigg]^{TT} \nonumber \\
&\mathcal{B}_{+}  
    = \frac{\pi\,  \alpha}{2}  \bigg[\frac{2}{v_r^2}+\frac{\left(1-v_r^2\right)} {v_r^3}\log
   \left(\frac{1-v_r}{1+v_r}\right)\bigg] =1.44 \pi \alpha \nonumber \\
& \mathcal{B}_{\times}=0
\end{align}
Incorporating the expression in Eq.~\eqref{plus_high_SI}, the final memory strain becomes
\be \label{eq:time_domain_memory_large_SI}
[h_{+}^{\text{mem}}(t, \vec x)]^{TT} = 1.438 \, \frac{G\alpha}{36 r} E_b \, (1-e^{-\frac{u-\sqrt{3}R_s}{\sqrt{3}\tau_\nu}}) \, \Theta(u-\sqrt{3}R_s). 
\ee
We find that as $u\to\infty$, $h^{mem}\sim 10^{-22}$. Thus, the signal in case of high $\nu$-SI is one order less than without SI.

\begin{figure}[htbp]
    \centering
    \includegraphics[scale=0.25]{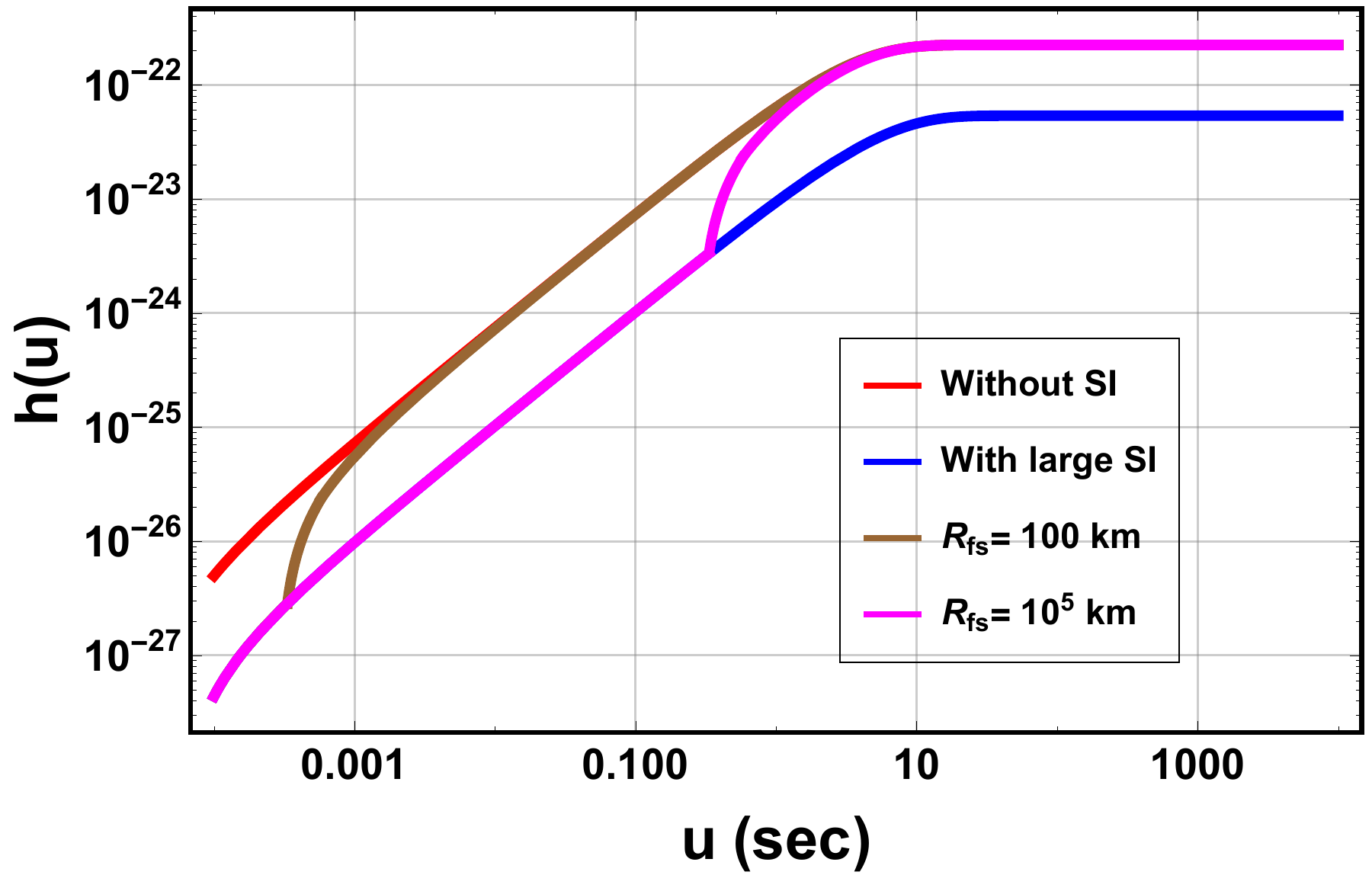}
    \caption{Log-log-plot for the time domain memory strain waveforms. The four waveforms correspond to neutrino propagation without self-interaction ({\em red}), with high self-interaction ({\em blue}), mixed neutrino propagation with $R_{fs}=100$ km ({\em brown}), and $R_{fs}=10^5$ km ({\em purple}). The plot reveals the build-up of the memory strain along $u$. We find that the amplitude for memory with $\nu$-SI is lower compared to the case when there is no SI ,{\em i.e.} when the neutrino free-streams. In cases where there is mixed propagation ($v=1/\sqrt{3}$ when $\nu$-SI is present and $v=1$ in the free streaming region), the waveforms show a transition corresponding to the velocity of the neutrino. With increase in the value of $R_{fs}$, the transition happens at a later $u$-value. The final memory value in all the mixed propagation scenarios is similar to the case when there is no $\nu$-SI. In the plots $\alpha=0.005$ and $R_s=10$km.}
    \label{fig:time_domain_memory}
\end{figure}

   When $\nu$-SI is moderate then both the integrals in Eq.~\ref{hijtgw-5a} needs to be computed. The final $h_{\rm mem} =h_{\rm mem}^I+h_{\rm mem}^{II}$ will have two contributions corresponding to the two regions that the neutrino passes through. Region I is where the velocity is $v_r=1/\sqrt{3}$ and region II is where it free streams with a speed of unity. The final form for the expressions are given below.
\begin{align}
    h_{+}^I = & 1.438 \, \frac{G\alpha}{36 r} E_b \,\, \Theta (u-{\sqrt{3}}R_s)[1-e^{-\frac{u-\sqrt{3}R_s}{\sqrt{3} \tau _{\nu }}}
    \nonumber\\
    & \hskip 1cm - \Theta (u-{\sqrt{3}}R_{fs})\, (1-e^{-\frac{u-\sqrt{3}R_{fs}}{\sqrt{3} \tau _{\nu }}})], \label{eq:reg_I_mixed_time}
   \\
     h_{+}^{II} = & \frac{G\alpha}{6 r} E_b \, (1-e^{-\frac{u- R_{fs}}{\tau_\nu}}) \, \Theta(u-R_{fs}). \label{eq:reg_II_mixed_time}
\end{align}
The Heaviside theta functions in the GW strain waveform signify transition from diffusion region to the free-streaming one. As is evident from Fig.(\ref{fig:time_domain_memory}), there is a transition when the neutrino value changes from $1/\sqrt{3}$ to unity. The transition takes at a later value of $u$ with the increase in the value of $R_{fs}$. This is because a higher value of $R_{fs}$ denotes that the neutrino spends more time in the diffusion region. In all the plots we find the memory rise time in order of 10s. This confirms that detectors like DECIGO, BBO are well-suited to detect this effect as the maximum characteristic strain $h_c(f)$ will peak around $\mathcal{O}(10^{-1})$Hz.

Similar scenarios can be studied with a smooth profile of neutrino radial velocity, such as, $v_r=\sqrt{1-(2 \,R_s^2)/(3\,r^2)}$.  In this profile, we find that at $r=R_S,v_r=1/\sqrt{3}$, but at large distances, $v_r=1$. Thus, asymptotically the neutrino free streams. We obtain numerically the strain amplitude, $h^{\text{mem}}\sim 4.88\times 10^{-21}$. \\

\noindent \underline{\em Detection prospects: }
Finally we discuss the possibility  of detecting memory signals with $\nu$-SI. In order to achieve this, we first compute the waveforms in frequency domain and then calculate the characteristic strain for such kind of burst profiles. The frequency domain waveforms are used to compare it with the power spectral density (PSD) corresponding to the different detectors. This enables us to understand the possibility of detecting a given signal in the corresponding detector. In frequency space, the waveform is given by the expression,
\be \label{eq:fourier_transform}
|\Tilde{h}(f)|=\int h(u) e^{2\pi i f u} \, du.
\ee

Incorporating the time domain waveforms from Eqs.~\eqref{eq:time_domain_memory_without_SI}, \eqref{eq:time_domain_memory_large_SI}, \eqref{eq:reg_I_mixed_time} and \eqref{eq:reg_II_mixed_time} in Eq.~\eqref{eq:fourier_transform}, we find closed form expressions for the frequency memory signals too. We enlist them below. 
 \begin{align}
     |\Tilde{h}_1(f)|=& \frac{G\alpha}{6 r} E_b \Theta (R_s\,^{-1}-2\pi  f) 
     \nonumber \\
    & \bigg|\pi \delta (2\pi f)+ \bigg(\frac{i}{2\pi f}+\frac{\tau_\nu}{-1+2\pi i f \tau_\nu}\bigg)e^{2\pi i f R_s}\bigg|, \, \, \label{eq:freq_memory_without_SI}\\
      |\Tilde{h}_2(f)| = & 0.04 \, \frac{G\alpha}{r} E_b\, \Theta ((\sqrt{3}R_s)\,^{-1}-2\pi  f) 
      \nonumber \\
      & \bigg|\pi \delta (2\pi f)+ \bigg(\frac{i}{2\pi f}+\frac{\sqrt{3}\, \tau_\nu}{-1+2\pi i f \sqrt{3}\, \tau_\nu}\bigg) e^{2\pi i f \sqrt{3} \, R_s}\bigg|.  \label{eq:freq_memory_large_SI}
 \end{align}
 Eqs.~\eqref{eq:freq_memory_without_SI} and \eqref{eq:freq_memory_large_SI} denote the frequency space memory waveforms for without $\nu$-SI and large $\nu$-SI, respectively. The corresponding Heaviside theta functions in frequency space follows from restriction imposed in the time domain, {\em viz.} $u>R_s$ in without $\nu$-SI and $u>\sqrt{3} R_s$ for large $\nu$-SI.  In the final case the neutrino travels (in the diffusion region) from $R_s$ to $R_{fs}$ with $v=1/\sqrt{3}$ and for $r>R_{fs}$, it free streams. This is the mixed propagation mode where the $\nu$-SI is moderate. 
 \begin{widetext}
    \begin{eqnarray} 
     |\Tilde{h}_3(f)| = && 1.438 \, \frac{G\alpha}{36 r} E_b \bigg|\pi \delta (2\pi f)+ \bigg(\frac{i}{2\pi f}+\frac{\sqrt{3}\, \tau_\nu}{-1+2\pi i f \sqrt{3}\, \tau_\nu}\bigg) \text{Exp}[2\pi i f \sqrt{3} \, R_{s}]\bigg|  \, \, \Theta ((\sqrt{3}R_{s})\,^{-1}-2\pi  f) \nonumber \\
     && - 1.438 \, \frac{G\alpha}{36 r} E_b \bigg|\pi \delta (2\pi f)+ \bigg(\frac{i}{2\pi f}+\frac{\sqrt{3}\, \tau_\nu}{-1+2\pi i f \sqrt{3}\, \tau_\nu}\bigg) \text{Exp}[2\pi i f \sqrt{3} \, R_{fs}]\bigg|  \, \, \Theta ((\sqrt{3}R_{fs})\,^{-1}-2\pi  f)  \nonumber \\
     && + \frac{G\alpha}{6 r} E_b \bigg|\pi \delta (2\pi f)+ \bigg(\frac{i}{2\pi f}+\frac{\tau_\nu}{-1+2\pi i f \tau_\nu}\bigg) \text{Exp}[2\pi i f R_{fs}]\bigg| \Theta (R_{fs}\,^{-1}-2\pi  f). \label{eq:freq_memory_mixed}
 \end{eqnarray} 
 \end{widetext}

 \begin{figure}[htbp]
    \centering
    \includegraphics[scale=0.25]{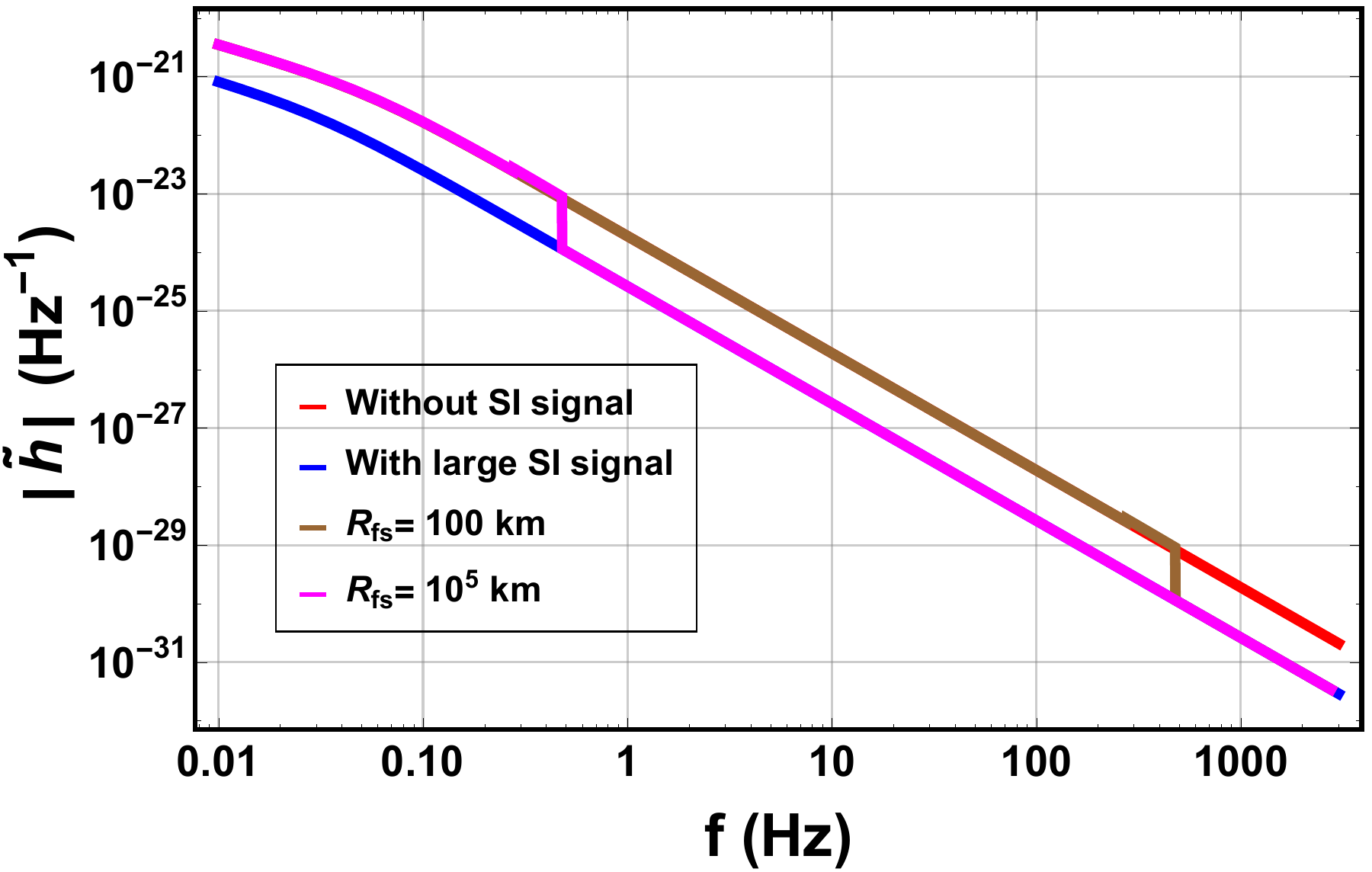}
    \caption{Log-log plot for the frequency domain memory waveforms. These profiles are generated by the fourier transform of the time-domain waveforms. We observe that with the rise in frequency, the velocity of the neutrino in mixed propagation scenarios, becomes equal to $v=1/\sqrt{3}$. The transition happens later for lower values of $R_{fs}$. This is expected since in the time-domain waveforms, a lower value of $R_{fs}$ has a earlier transition.}
    \label{fig:freq_domain_waveform}
\end{figure}
Frequency domain waveforms are shown in all the these cases in Fig.~\eqref{fig:freq_domain_waveform}. We observe that similar transition in frequency-space too. With higher values of $R_{fs}$, the diffusion region increases, and the mixed propagation mode transits earlier in frequency space from $v=1$ mode to $v=1/\sqrt{3}$.  This is consistent with the results obtained in the time-domain waveforms. {\em This transition feature for mixed neutrino propagation mode is indicator of the $\nu$-SI.}

\begin{figure}[htbp]
    \centering
    \includegraphics[scale=0.25]{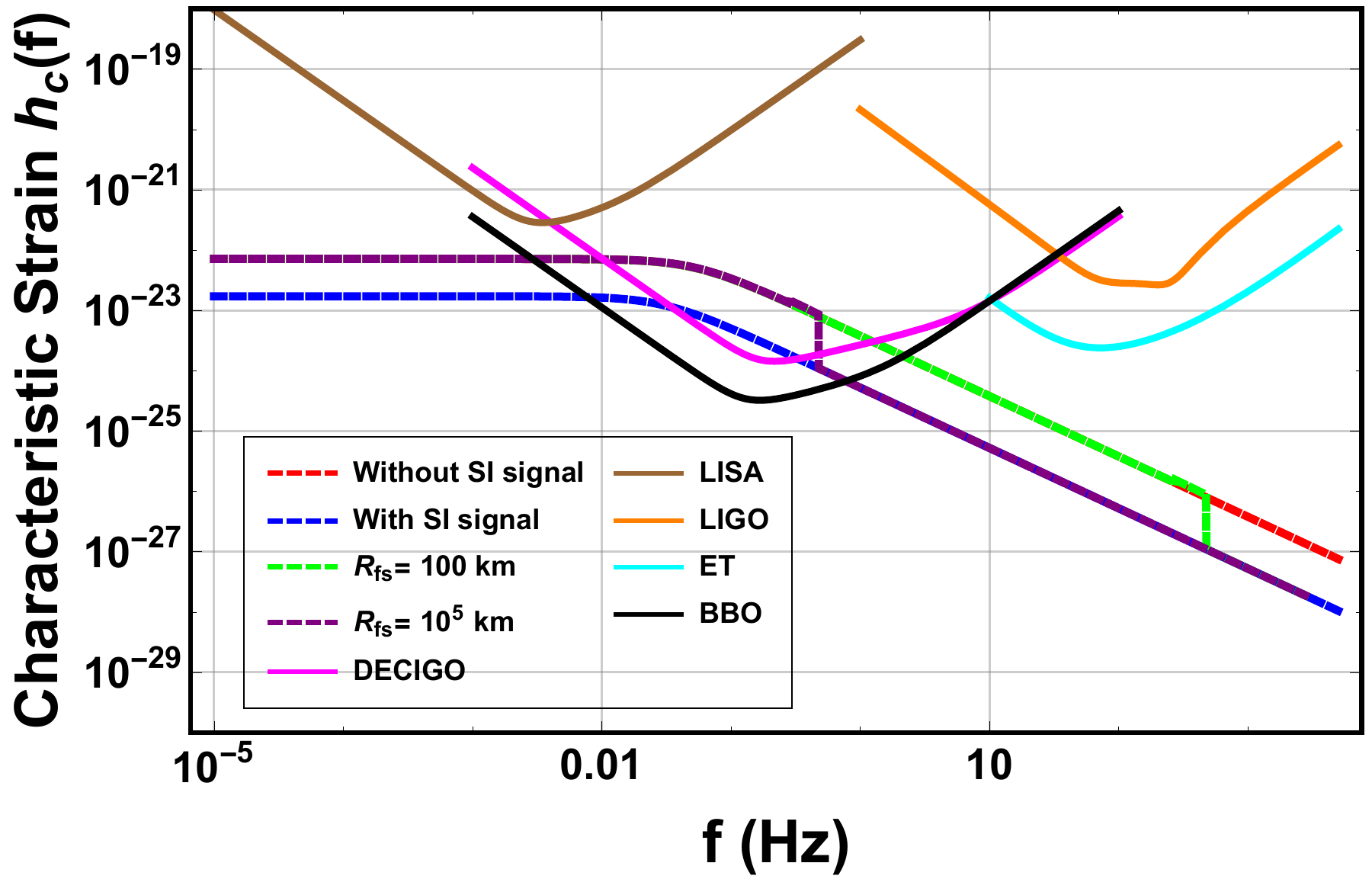}
    \caption{Characteristic strain for the memory signal is plotted along with the characteristic noise of the detectors. The plots show that only DECIGO and BBO are able to detect the memory signal. The transition occurring in $R_{fs}=10^5$km can also be detected by BBO and DECIGO. For lower values of $R_{fs}$, {\em i.e.} smaller diffusion region, the transition occurs at a higher frequency value which is beyond the sensitivity of the current and future detectors.}
    \label{fig:detectability}
\end{figure}

Next we compute the characteristic strain and (signal-to-noise ratio) SNR values corresponding to current and upcoming GW detectors. The characteristic strain amplitude $h_c(f)$ and its noise counterpart $h_n(f)$ are,
\begin{gather} \label{eq:ch_strain}
    [h_c\, (f)]^2=4 f^2  |\Tilde{h}(f)|^2 \hspace{2cm} 
    [h_n(f)]^2=f S_n(f)
\end{gather}
 $S_n(f)$ denotes the PSD of a detector. We obtain the PSD for detectors like aLIGO, ET, LISA from \cite{Sathyaprakash:2009} and, for DECIGO and BBO from \cite{Yagi:2011}. Finally, the SNR for memory signals involving $\nu$-SI are enlisted in Table-1. 
    \begin{gather}
    (SNR)^2=\int_{-\infty}^{\infty} d(\text{log}\, f)\, \frac{[h_c\, (f)]^2}{ [h_n(f)]^2} \label{eq:charactertic_strain}
\end{gather}

In Fig.~\eqref{fig:detectability}, we analyse the observational potential of the memory signals in some current and upcoming detectors. LIGO, ET, LISA sensitivity curves lie above the characteristic strain of the memory and hence are unable to detect this signal. Only DECIGO and BBO are well-suited to observe this effect of $\nu$-SI. Moreover, we find initially $h_c(f)$ is a constant which corresponds to the {\em zero-frequency limit} \cite{Mukhopadhyay:2021zbt, Kolb:1987, Epstein:1978}. The transition for lower values of $R_{fs}=100$ km happens at a higher frequency which is beyond the detectability regime of any of these detectors.  Nevertheless, we find that for higher values of $R_{fs}$, there is significant potential to detect this signal. The SNR values quoted in Table-I show that detecting the signal without $\nu$-SI and with $\nu$-SI for $R_{fs}$ is similar. For large $\nu$-SI the SNR drops significantly. Thus, as brought out from the analysis, we conclude that the detectors DECIGO and BBO may as well detect this signal since they operate in the range $0.1-10$ Hz. {  For lower values of $R_{fs}$, the signals are indistinguisable from the without $\nu$-SI memory signal. This corresponds to the case when $\nu$-SI strength $G^\prime$ becomes identical to the weak interaction strength $G_F$.}

\begin{center}
\begin{table}
\begin{tabular}{ |c|c|c| } \hline
 \multicolumn{3}{|c|}{SNR values}\\
 \hline
 Memory profiles & DECIGO & BBO \\ 
 \hline
 Without SI ($v=1$)  & 3.65 & 7.88 \\ 
 \hline
 With SI ($v=1/\sqrt{3}$)  & 1.40 & 3.06 \\ 
 \hline
 $R_{fs}= 100$ km   & 3.65 & 7.88 \\
  \hline
  $R_{fs}= 10^5$ km  & 3.36 & 7.46 \\
  \hline
\end{tabular}
\caption{The SNR corresponding to the memory profiles for each of the five detectors are noted.}
\end{table}
\end{center}
\noindent \underline {\em Conclusions:} In this letter, we have tried to showcase for the first time how secret $\nu$-SI can lead to significant changes in the gravitational memory profile for a  supernova neutrino burst. To this end, we have computed the memory waveforms in both time and frequency domain and have explicitly shown that there exists possibility of detecting such signals using detectors like BBO and DECIGO, thereby enabling our claim.

The basic premise of this work deals with SI of Majorana neutrinos in the Majoron model. In this model, a mediator scalar field having masses of the order of $10$ MeV, gives rise to an effective interaction which is higher compared to the weak-interaction scale. Since the coupling is high, the neutrinos coming out of PNS, have smaller mean free paths and, hence, are unable to free-stream. In this region, neutrinos travel as pressure waves with velocity ($v=1/\sqrt{3}$). As the density falls, the neutrino starts free-streaming from $R_{fs}$. A higher value of $R_{fs}$ implies larger diffusion region, leading to stronger interaction and thereby, smaller mean free-path. In our entire analysis, we work with two values of $R_{fs}$, {\em i.e.}, $100\,$km and $10^5\,$km.  {  The values for $\nu$-SI coupling is provided for $R_{fs}=100$ km and $10^5$ km, respectively, with two distinct cases of power law fall-offs of the neutrino density.}

We have obtained closed-form expressions for both time-domain and frequency domain memory waveforms. The waveforms are obtained for neutrino burst models with constant anisotropy parameter. In order to ascertain the role of $\nu$-SI vividly, we also analyse two opposite cases-- i) where there is no $\nu$-SI; $R_{fs}=R_s, v_r=1$, ii) when the $\nu$-SI is large; $R_{fs}\to \infty, v_r=1/\sqrt{3}$.  We find that in the time domain, a lesser vale of $R_{fs}$ shows transition earlier from large $\nu$-SI to without $\nu-$SI case. This is because the neutrinos spend less time in the diffusion zone when $R_{fs}$ is small. The frequncy plots, expectedly, shows the opposite behaviour (Fig.(\ref{fig:freq_domain_waveform})). One thing to note is that the transitions in the figures are steep  due to the discontinuous nature of the velocity profile we have chosen. In case of smooth velocity profile, the transition will also be smooth. But, the overall feature of the memory waveforms will remain the same. 

We find that the detectability of such $\nu$-SI memory signals is achievable with planned space-borne detectors DECIGO and BBO. The transition is observable for large value of $R_{fs}$. For smaller values like $R_{fs}=100 \,$km, the signal is indistinguishable from the free-streaming scenario. We require higher detector sensitivities at kHz frequency range. 

Finally, to conclude, gravitational wave astronomy holds promise to probe fundamental physics in the strong gravity regime. A more challenging work will be to consider realistic burst models (some of them are given in \cite{Mukhopadhyay:2021zbt}) and find out the features of this memory signal in those cases. We hope to address these issues in future.  
\\

\noindent\underline{\em Acknowledgements:} {  The authors thank the anonymous referee for the insightful comments and also suggesting changes which led to the overall improvement in the quality of the letter.} I. C. thanks Praveer Tiwari and Sayantan Ghosh for discussions regarding the detectability of the proposed signal. He also gratefully acknowledges Indian Institute of Technology Bombay for providing financial assistance through postdoctoral fellowship (Employee code: $20003307$). S. B. acknowledges D. A. E.  for providing a post-doctoral fellowship (grant no: $1003/(6)/2021/\text{RRF/R\&D-II}/4031$). A. H. would like to thank the M.H.R.D., Government of India for the research fellowship.

%\section{Microphysics: Weak interaction strength}

%The detection of self-interactions has also interesting implications for fundamental physics, in relation to the strength of weak interaction. In \cite{Chang:2022aas}, an expression is given relating the strength of neutrino self interaction with $R_{fs}$. The numbers show that for $R_{fs}=10 \, km$, the modified Fermi coupling $G'\sim 10^{-22}\, (MeV)^{-2}$ and with $R_{fs}=10^5 \,km, G'\sim 10^{-14} \, (MeV)^{-2}$. This simply shows that higher the value of $R_{fs}$, larger is the diffusion range, leading to stronger interaction and thereby, smaller mean free-path. Thus, small $\nu$-SI can be detected in this scenarios. 

\appendix

\section{Gravitational waves from relativistic fluids}
\label{app:GW_formula} 

\subsection{Relativistic point particles}

The stress tensor for massive particles is given by
\be
T^{ij}(t^\prime, \vec x^\prime)= \sum_a \gamma_a m_a v_a^i v_b^i \delta^3(\vec x^\prime- \vec x_a(t^\prime))
\label{Tij}
\ee
where $v_a^i$ is the velocity of the particle labelled '$a$' and $\gamma_a=(1-v_a^2){-1/2}$ is the Lorentz factor.  Here $\vec x_a(t^\prime)$ describes the trajectory of the source particle '$a$'.

The  gravitational waves from the sourced by a stress tensor $T^{ij}$  obey the wave equation
\begin{eqnarray}
\square \,h_{ij}(t, \vec x)=- 16 \pi G\, T_{ij}(t, \vec x).
\label{sourced_wave_eqn}
\end{eqnarray}
The solution of Eq.~\eqref{sourced_wave_eqn} is of the form
\begin{eqnarray}\label{hijtgw-1}
h^{ij}(t, \vec x)= 4G \int dt^\prime \,d^3 x^\prime\,\, T^{ij} (t^\prime, \vec x^\prime) \, \cfrac{\delta\left(t^\prime -(t-|\vec x -\vec x^\prime|)\right)}{|\vec x -\vec x^\prime|}.
\nn \\
\end{eqnarray}
where $\vec x^\prime$ is the point where the graviton is emitted and $\vec x$ is the location of the observer. Substituting from Eq.(\ref{Tij}) we obtain
\begin{align}\label{hijtgw-2}
h^{ij}(t, \vec x)=& 4G \int dt^\prime \,d^3 x^\prime\,\cfrac{\delta\left(t^\prime -(t-|\vec x -\vec x^\prime|)\right)}{|\vec x -\vec x^\prime|}
\nonumber\\ 
& \hskip 1cm \sum_a \gamma_a m_a v_a^i v_b^i \delta^3(\vec x- \vec x_a(t^\prime)) 
 \nonumber\\ 
=&4G \int dt^\prime \,  \sum_a \gamma_a m_a v_a^i v_b^i \, \cfrac{\delta\left(t^\prime -(t-|\vec x -\vec x_a(t^\prime)|)\right)}{|\vec x -\vec x_a(t^\prime)|}.
\end{align}

The distance of the observer is much larger than the source size,  $|\vec x| \gg |\vec x_a(t^\prime)|$ , take the approximations
\begin{eqnarray}
 \delta\left(t^\prime -(t-\lvert\vec x -\vec x_a(t^\prime)\rvert)\right) &\simeq& \delta(t^\prime- (t-r+\vec N\cdot \vec x_a(t^\prime)))\nn\\
 \nn\\
 &=& \frac{\delta \left(t^\prime- (t-r) \right)}{\frac{d}{dt^\prime}(t^\prime- (t-r+\vec N\cdot \vec x_a(t^\prime)))}
 \nn\\
  &=& \frac{\delta \left(t^\prime- (t-r) \right)}{ 1-\vec N \cdot \vec v_a(t^\prime)} \,.
\end{eqnarray}
where $\vec x \equiv r \vec N$.
We also take
$\cfrac{1}{|\vec x-\vec x_a(t^\prime)|}\simeq \cfrac{1}{r}$ and with these approximations  after performing the $t^\prime$ integral using the delta function, Eq.(\ref{hijtgw-2}) reduces to the form
\be
h^{ij}(t, \vec x)= \frac{4G}{r}  \sum_a \gamma_a m_a  \frac{v_a^i(t_r) v_a^j(t_r)}{ 1-\vec N \cdot \vec v_a(t_r)}
\ee
where $t_r=t-r$.

\subsection{Relativistic fluid}
For a macroscopic number of particles which source GRW, we go to the fluid limit of the stress tensor which is given by
\be
T^{ij}(t^\prime, \vec x^\prime) = \epsilon(t^\prime, \vec x^\prime) v^i(t^\prime, \vec x^\prime) v^j(t^\prime, \vec x^\prime)
\ee
where $\epsilon((t^\prime, \vec x^\prime)$ is the energy density and $\vec  v(t^\prime, \vec x^\prime)$ the velocity of the fluid element at the spatial location $\vec x^\prime( t^\prime)$.
A similar derivation as in the previous sub-section gives us the gravitational wave signal in terms of the source as
\begin{eqnarray}\label{hijtgw-3}
\hspace*{-0.3in} h^{ij}(t, \vec x)
%&=& 4G \int dt^\prime \,d^3 x^\prime\,\,  \sum_a \gamma_a m_a v_a^i v_b^i \delta^3(\vec x- \vec x_a(t^\prime)) \, \cfrac{\delta\left(t^\prime -(t-|\vec x -\vec x^\prime|)\right)}{|\vec x -\vec x^\prime|}. \nn\\
=\frac{4G}{r} \int dt^\prime \, d^3 x^\prime \epsilon(\vec x^\prime, t^\prime) \frac{v^i(\vec x^\prime, t^\prime) v^j(\vec x^\prime, t^\prime)}{ 1-\vec N \cdot \vec v (\vec x^\prime, t^\prime) } \nn\\
\delta \left(t^\prime- (t-r) \right).
\end{eqnarray}
We can see that Eq.(\ref{hijtgw-3}) can be derived from Eq.(\ref{hijtgw-2}) by 
making  the following replacement in going to the fluid limit
\be
\sum_a \gamma_a m_a  \Rightarrow \int d^3 x^\prime \epsilon(t^\prime, \vec x^\prime)\,\quad {\rm and} \quad v_a^i(t^\prime) \Rightarrow v^i (\vec x^\prime(t^\prime))\,.
\ee

\subsection{Null fluids}
For fluids which move at the speed of light with trajectories given by null-geodesics the stress tensor components can be written as
\be
T^{ij}(t^\prime, \vec x^\prime) = T^{00} (t^\prime, \vec x^\prime) {n^\prime}^i(t^\prime, \vec x^\prime) {n^\prime}^j(t^\prime, \vec x^\prime)
\ee
here ${n^\prime}^i(t^\prime, \vec x^\prime)= v^i/|\vec v|$. With $|\vec v|=c=1$, ${n^\prime}^i$  are components of the unit vector. For a radial flux of mass-less particles as the case of neutrinos from supernova ${n^\prime}^i=\frac{{x^\prime}^i}{|\vec x^\prime|} =(\sin \theta^\prime \cos \phi^\prime, \sin \theta^\prime \sin \phi^\prime, \cos \theta^\prime)$.

The gravitational signal from radially radiated null-fluids have the form 
\begin{eqnarray}\label{hijtgw-4}
h^{ij}(t, \vec x)
%&=& 4G \int dt^\prime \,d^3 x^\prime\,\,  \sum_a \gamma_a m_a v_a^i v_b^i \delta^3(\vec x- \vec x_a(t^\prime)) \, \cfrac{\delta\left(t^\prime -(t-|\vec x -\vec x^\prime|)\right)}{|\vec x -\vec x^\prime|}. \nn\\
=\frac{4G}{r} \int dt^\prime \, d^3 x^\prime \epsilon(\vec x^\prime, t^\prime) \frac{{n^\prime}^i {n^\prime}^j}{ 1-\vec N \cdot \vec  {n^\prime}} \delta \left(t^\prime- (t-r) \right).
\nn\\
\end{eqnarray}
where $\epsilon(\vec x^\prime, t^\prime)=T^{00}(\vec x^\prime, t^\prime)$ is the energy density of the null fluid.

{   
\section{Spatial dependence of the anisotropy} \label{app:rad-ani}

Considering the radial dependence of the anisotropy parameter can lead to significant changes in the gravitational memory signal. We outline a few cases below and provide qualitative comments on how the gravitational wave (GW) signal is modified compared to the scenario without spatial dependence, as discussed in the main letter. 

\begin{enumerate}

    \item \underline{$\alpha (r,\theta,\phi)= \alpha_0 \, \sin^2\phi\, e^{-r/Rs}$:} \\
    We assume the anisotropy follows the functional form described above throughout the region from $R_s<r<\infty$. Now, at $r=R_{fs}$, the anisotropy becomes, 
    \begin{equation}
        \alpha (\theta,\phi) =\alpha_0 \, \sin^2\phi\, e^{-R_{fs}/Rs}
    \end{equation}
If $R_{fs}>>R_s$, as in the case of $R_{fs}=10^5$ km, then the anisotropy goes to zero leading to vanishing memory signal from the second term in Eq.(13) of the main letter. The integral in the diffusion region also gives much fainter memory signal owing to the same exponential factor of $e^{-r/Rs}$ as $r$ goes from $R_s \, \text{to} \, R_{fs}$.
     
     \item \underline{$\alpha (r,\theta,\phi)= \alpha_0 \, \sin^2\phi\, e^{-r/R_{fs}}$:}\\
Similar to earlier case, we find that the second term in Eq.(13) of the main letter to have negligible constibution as $r>>R_{fs}$. Since $r=10$ kpc is quite large than the free-streaming radius (both for $R_{fs}=100$ km and $10^5$ km),  the only contribution comes from the diffusion region. 

\item \underline{$\alpha (r,\theta,\phi)= \alpha_0 \, \sin^2\phi \,   (1-e^{- r/R_{fs}\, } )\,  $:}\\
Assuming such a profile, we find two limiting cases. When $r<<R_{fs}$, then 
\begin{equation}
    \alpha (r,\theta,\phi)= \alpha_0\, \sin^2\phi\, \frac{r}{R_{fs}}. 
\end{equation}
This shows that near $R_s$, the GW memory signal in this case will be less than the corresponding obtained without any radial dependence due to the factor of $r/R_{fs}$. On the other limit, $r>>R_{fs}$, we get back the same anisotropy profile as worked out in the main letter,
\begin{equation}
    \alpha (r,\theta,\phi) = \alpha_0\, \sin^2\phi.
\end{equation}

\begin{figure}[htbp]
    \centering
    \includegraphics[scale=0.35]{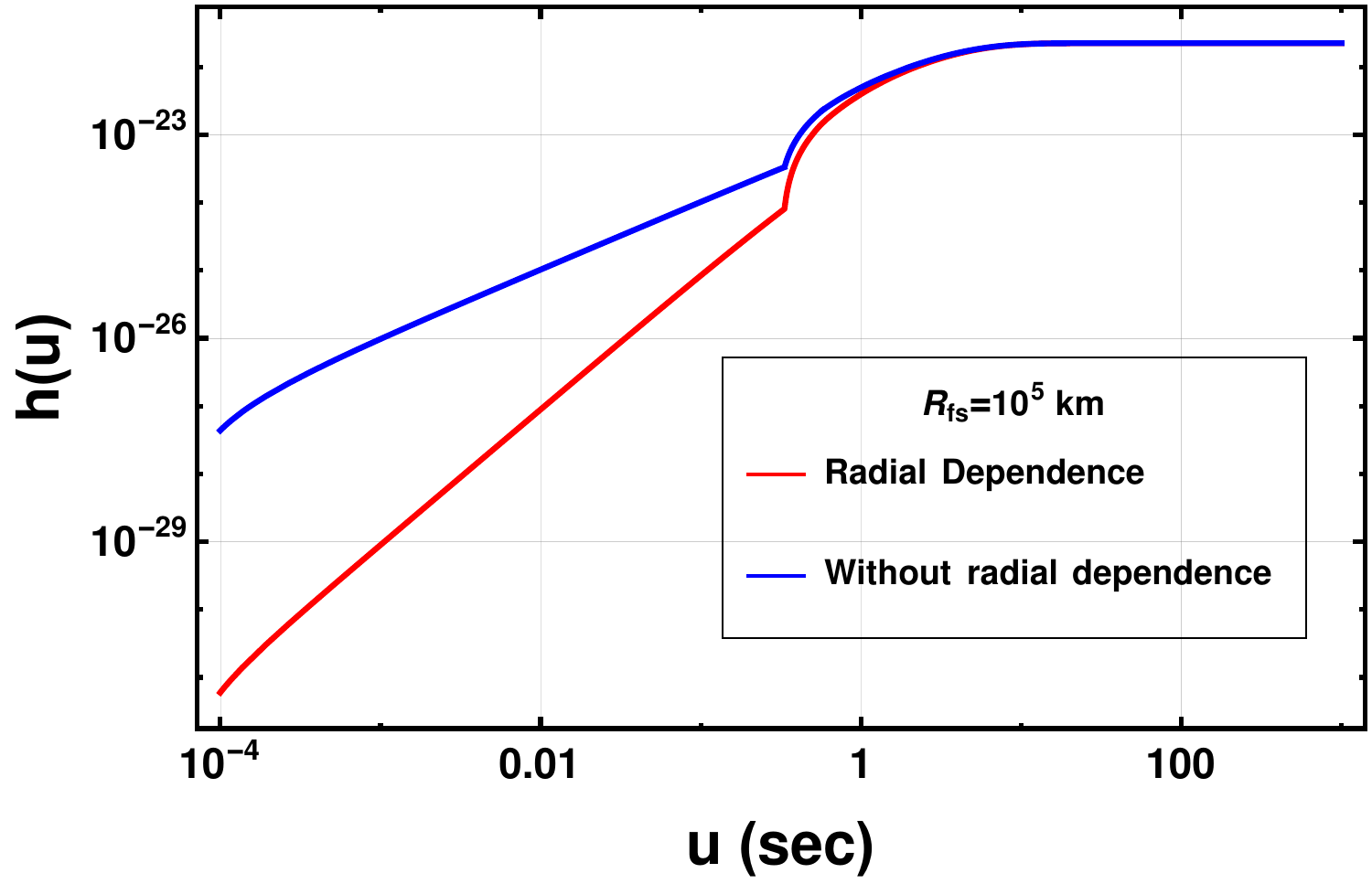}
    \caption{\small Log-log-plot for the time domain memory strain waveforms with radial dependence given by $ \alpha=\alpha_0  \sin^2\phi (1-e^{- r/R_{fs}\, } ) $ in red and $\alpha=\alpha_0  \sin^2\phi $ in blue. The two cases differ in the diffusion region ($R_S<r<R_{fs}$) but they tend to show similar values as $r>R_{fs}$. In both cases, $\alpha_0=0.005$ as considered in the main letter.}
    \label{fig:rad-dep}
\end{figure}

In Fig.(\ref{fig:rad-dep}), We have demonstrated how anisotropy behavior modifies the GW memory time-domain waveform. Initially, we find that the case involving radial dependence of anisotropy results in a lesser GW strain compared to the case without radial dependence of anisotropy. However, after $R_{fs}$, the two strains seem to merge.

\end{enumerate}

\bibliographystyle{apsrev4-2.bst}
\bibliography{bibliography}

\end{document}